\documentclass[conference]{IEEEtran}
\IEEEoverridecommandlockouts
\usepackage{arabtex}
\usepackage{utf8}
\setcode{utf8}
\usepackage{cite}
\usepackage{amsmath,amssymb,amsfonts}
\usepackage{algorithmic}
\usepackage{graphicx}
\usepackage{textcomp}
\usepackage{xcolor}

\usepackage{times}  
\usepackage{helvet}  
\usepackage{courier}  
\usepackage[hyphens]{url}  
\usepackage{graphicx} 
\usepackage{caption} 
\usepackage{algorithm}
\usepackage{algorithmic}

\usepackage{booktabs}

\newcommand{\pb}[1]{\vspace{0.75ex}\noindent{\bf \em #1}\hspace*{.3em}}

\def\BibTeX{{\rm B\kern-.05em{\sc i\kern-.025em b}\kern-.08em
    T\kern-.1667em\lower.7ex\hbox{E}\kern-.125emX}}
\begin{document}

\title{A Twitter Dataset for Pakistani Political Discourse}

\author{
\IEEEauthorblockN{
Ehsan-Ul Haq\IEEEauthorrefmark{1}, 
Haris Bin Zia\IEEEauthorrefmark{2},
Reza Hadi Mogavi\IEEEauthorrefmark{1}, 
Gareth Tyson\IEEEauthorrefmark{3}, 
Yang K. Lu \IEEEauthorrefmark{1},
Tristan Braud\IEEEauthorrefmark{1}, 
and Pan Hui\IEEEauthorrefmark{1}\IEEEauthorrefmark{3}}
\IEEEauthorblockA{\IEEEauthorrefmark{1}Hong Kong University of Science and Technology, HKSAR}
\IEEEauthorblockA{\IEEEauthorrefmark{2}Queen Mary University of London, UK}
\IEEEauthorblockA{\IEEEauthorrefmark{3}Hong Kong University of Science and Technology, Guangzhou}
Email:     \{euhaq, rhadimogavi\}@connect.ust.hk \quad \{h.b.zia\}@qmul.ac.uk \quad \{gtyson, yanglu, braudt, panhui\}@ust.hk \\
}

\maketitle

\begin{abstract}
We share the largest dataset for the Pakistani Twittersphere consisting of over 49 million tweets, collected during one of the most politically active periods in the country. We collect the data after the deposition of the government by a No Confidence Vote in April 2022. This large-scale dataset can be used for several downstream tasks such as political bias, bots detection, trolling behavior, (dis)misinformation, and censorship related to Pakistani Twitter users. In addition, this dataset provides a large collection of tweets in Urdu and Roman Urdu that can be used for optimizing language processing tasks.
\end{abstract}

\section{Introduction}

\noindent Globally, Twitter is one of the leading data sources related to political and social communication studies~\cite{murthy_twitter_2013,lorenz-spreen_systematic_2022} and Twitter datasets have been used for several tasks related to computational politics~\cite{haq_survey_2020-1}, polarization~\cite{budhiraja2021american}, propaganda, and political trolling~\cite{alhazbi2020behavior}. Most of the publically shared dataset and research is heavily focused on western countries, such as the US. However, Pakistan, where Twitter is one of the primary sources for political discourse, remains largely an under-studied population in research related to social network analysis and social computing~\cite{hussain_analyzing_2021,ul_haq_enemy_2020}.

Since the 2013 general elections in Pakistan, Twitter has been heavily utilized by major political parties in Pakistan~\cite{ahmed_twitter_2015,ahmed_my_2014}. There are reports of targeted campaigns, propaganda, and fake accounts deployed by political parties to get political advantage~\cite{mir_political_2022}. Particularly, the utilization of social media in the 2018 elections for political engineering. Winner of the 2018 elections, the government of Pakistan Tehreek Insaf (PTI) was also criticized for using social media campaigns for maligning other political parties and running state-funded campaigns through government media cells~\cite{noauthor_troubled_2013}. 

In April 2022, Pakistan went through one of the biggest political and constitutional crisis in the country~\cite{baloch_pakistan_2022,hrw_constitutional_2022}, when the prime minister was ousted from office through a no-confidence vote. This led to a large online discourse and activism based on several narratives from supporters of almost all political parties. This also was followed by a series of protests and processions across the country. This timely collection of the dataset can help several downstream research directions ranging from misinformation, propaganda, polarization, and natural language processing task for languages such as Urdu and Hindi.

\section{Dataset}

\subsection{Collection} 

We use Twitter streaming API to collect the data.\footnote{\url{https://developer.twitter.com/en/docs/tutorials/stream-tweets-in-real-time}} Streaming API collects the data in real-time and is a standard API for collecting Twitter dataset. We use the name of political leaders in the country and trending hashtags related to political narratives of different political parties to collect the data. The keywords are shown in Table~\ref{tab:keywords_list}. The data is collected from 19\textsuperscript{th} to 7\textsuperscript{th} of May for a total of 17 days. 
We do not apply additional filters such as location or language, hence the dataset consists of global discourse. 

\begin{figure}[t]
    \centering
    \includegraphics[width=.37\textwidth]{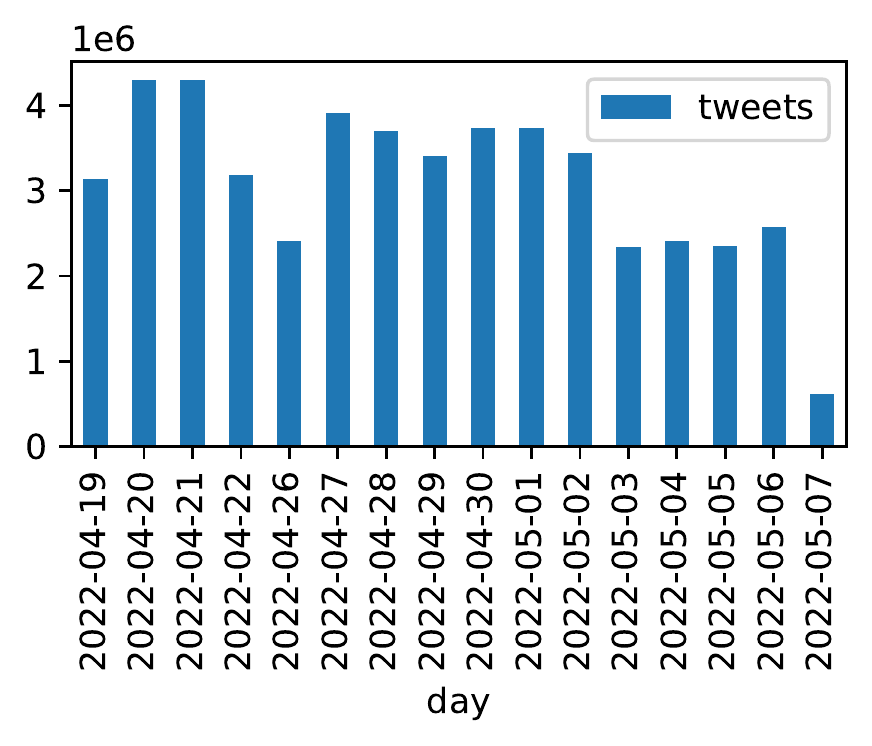}
    \caption{Total Number of tweets (in millions) per day.}
    \label{fig:tweet_per_day}
\end{figure}

Our dataset is available at Zenodo DOI\footnote{\url{https://doi.org/10.5281/zenodo.7538667}} \cite{haq_twitter_2023_a} and consists of 49,549,545 tweets collected over 17 days. The distribution of tweets is shown in Figure~\ref{fig:tweet_per_day}.


\begin{table*}[t]
\centering
\begin{tabular}{llllll}
\toprule
pakistan & imran khan & pmik & pti & ppp & pppp \\
pmln & noconfidence & nawaz sharif & shahbaz sharif & bilawal & zardari \\
bhutto & maryam nawaz & imrankhan & \<مریم نواز> & \<عمران خان> & \<پاکستان> \\
\<زرداری> &\<بلاول> &\<نوازشریف> &\<امپورٹڈحکومت نامنظور> &\<شھباز شریف> &\<امپورٹڈحکومت> \\
\< وزیر۱عظم> & behindyouskipper &\<> &\<> &\<> &\<> \\
\midrule
\end{tabular}
\caption{Keyword list used for the dataset collection}
\label{tab:keywords_list}
\end{table*}


\subsection{Exploratory Analysis}
In this section, we report the aggregated statistics and metadata of the dataset.

\pb{Hashtags and Mentions}
The dataset contains 47,690 and 382,083 unique hashtags and mentions respectively. These hashtags and usernames have been used 18,743,337 and 44,997,458 times, respectively. 

In addition, we show the top 10 most used hashtags and accounts in the dataset in Table~\ref{tab:top_10_hts} and \ref{tab:top_10_mentions}, respectively. We note that among the most used hashtags, most are from the deposed government narrative where the focus is demand for an earlier general election\footnote{1st and 3rd hashtag differ on spaces used between the words}. There is one hashtag, particularly, that was mostly used by the newly formed government. Among the top-10 most mentioned accounts, the most mentioned account is of a journalist, followed by politicians accounts and a YouTube Channel \textit{Haqeeqat TV}. 

We note that one of the accounts with the screen name @GeneralWrites in the top 10 most mentioned accounts has been suspended. We also observe that this account was mostly mentioned together with pro-PTI accounts.

\pb{Language} We use the twitter assigned language codes to analyze the language of tweets. Overall, there are 61 languages attributed to all of the tweets. The most common languages are Urdu (the national language) and English (previously the official language). There are 34,588,431 tweets in Urdu, and 9,026,404 tweets in English. Another common writing style in the country is Roman Urdu.~\footnote{\url{https://en.wikipedia.org/wiki/Roman_Urdu}} Which is annotated by Twitter as \textit{hi} or \textit{in}. This way of writing is also common in India for the Hindi language. There are 3,484,425 tweets without any identified language.

\pb{Country of Tweet} Twitter API can provide geographical information for a tweeting user depending on the user's profile settings. However, only a smaller percentage of data usually contains this information. Based on the tweets that carry this information, we find that tweets have been posted from 129 countries. Other than Pakistan, most tweets came from the United Arab Emirates, the UK, Saudi Arabia, and the US. All of these countries host a significant number of Pakistani workers and immigrants. 

\begin{table}[t]
\centering
\begin{tabular}{ll}
\toprule
Hashtag & count \\
\midrule
\<امپورٹڈ حکومت نامنظور>&14,041,331 \\ 
MarchAgainstImportedGovt & 1,097,411 \\
\<امپورٹڈ  حکومت  نامنظور> & 338,023\\
\<الیکشن کراو پاکستان بچاو>&226,045 \\
LahoreJalsa&136,560\\
Pakistan&124,757\\
MarchAgainstlmportedGovt&104,191\\
ImranKhan&93,525\\
\<توہین مسجد نبوی نامنظور>& 88,074\\
PakistanNeedsElections&77,556\\
 \bottomrule
\end{tabular}
\caption{Top 10 hashtags in the dataset and their usage count}
\label{tab:top_10_hts}
\end{table}

\begin{table}[t]
\centering
\begin{tabular}{lll}
\toprule
\textbf{Account Name} & \textbf{Count} & \textbf{Type}\\
\midrule
ImranRiazKhan                    & 1,350,362 & Journalist                           \\
PTIofficial                      & 1,041,832 & Political Party                        \\
AnwarLodhi                       & 794,747  &  Journalist                         \\
fawadchaudhry                    & 784,784  & Politician                            \\
MaleehaHashmey                   & 760,200   & Journalist                          \\
SHABAZGIL                        & 687,629  & Politician                           \\
GeneralWrites                    & 509,128 & Account Suspended                            \\
QasimKhanSuri                    & 486,330  & Politician                           \\
SdqJaan                          & 485,986  & Politician                           \\
Haqeeqat\_TV                     & 484,160      & YouTube Channel \\
\bottomrule
\end{tabular}
\caption{Top 10 mentioned accounts}
\label{tab:top_10_mentions}
\end{table}

\section{Dataset Applicability}

\subsection{Media Bias}
Several news media outlets such as newspapers and TV channels are working in Pakistan.\footnote{\url{https://en.wikipedia.org/wiki/List_of_news_channels_in_Pakistan}}\textsuperscript{,}\footnote{\url{https://en.wikipedia.org/wiki/List_of_newspapers_in_Pakistan}} However, there is limited work on news media bias and partisan alignment of those media or social media users~\cite{qayyum_exploring_2018,ali_media_2019}. This dataset consists of tweets originating from several media outlets and journalists within the country and can be used in studies aiming to characterize the political bias of Pakistani news media. Such media bias studies can eventually help several other studies that rely on political bias measures and ground truth data for media sources~\cite{an_visualizing_2012,haq2022s}

\subsection{Censorship} 
\begin{figure}[t]
    \centering
    \includegraphics[width=.37\textwidth]{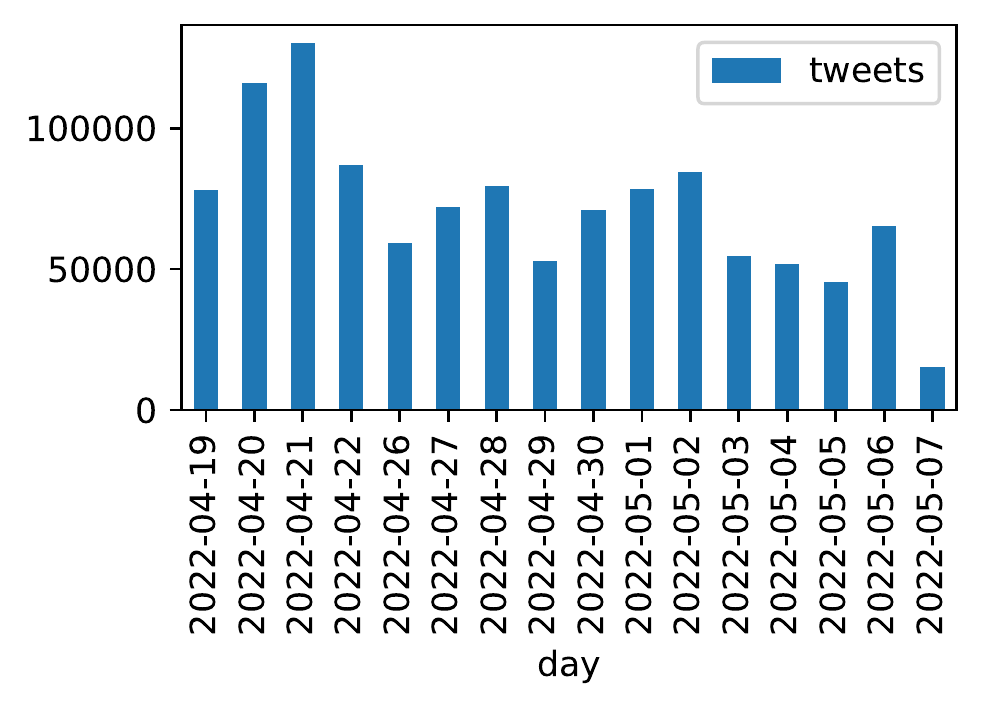}
    \caption{Total Number of censored tweets per day.}
    \label{fig:tweet_per_day_witheld}
\end{figure}

Political censorship is one of the concerning issues related to political discussions~\cite{elmas2021dataset}. Certain governments have been reported to censor political content, including Pakistan~\cite{nabi2013anatomy}. Our dataset contains a total of 950,208 tweets censored in 31 countries. We do not find any tweet in this dataset that has been censored in Pakistan. The largest number of tweets are censored in India, with India alone censoring 927,679 tweets. Given the historical relationship between Pakistan and India, such a large number is not surprising. However, these tweets can help researchers analyze this censorship from a large data perspective. We provide a list of top-5 countries that have censored the most tweets from this data in Table~\ref{tab:top_5_censored} and the daily distribution of censored tweets in Figure~\ref{fig:tweet_per_day_witheld}. 

This dataset can be used to study censorship practices across different countries. Particularly, large censorship from India can highlight several dimensions related to issues related to both countries, such as nationalism.

\begin{table}[t]
\centering
\begin{tabular}{ll}
\toprule
\multicolumn{1}{r}{\textbf{Country}} & \multicolumn{1}{r}{\textbf{Censored Tweets}} \\
\midrule
India                                & 927,679                             \\
Germany                                & 1,240                               \\
France                                & 799                                \\
Italy                                & 798                                \\
Sweeden                                & 798         \\
\bottomrule
\end{tabular}
\caption{Top 5 countries that censored tweets}
\label{tab:top_5_censored}
\end{table}

\subsection{Political Engineering}
Political Engineering through social media is a growing and global phenomenon~\cite{tufekci_engineering_2014,haq_survey_2020-1, haq_twitter_2022-1}. The ease of social media content generation and consumption leaves a common social media user susceptible to disinformation and astroturfing campaings~\cite{keller_political_2020}

Given the anecdotes of using social media for astroturfing and disinformation within the country, this dataset can help identify such practices, and also methods to effectively counter such practices. Even in our aggregated statistics, we note that one of the accounts among the top 10 most mentioned accounts is already suspended. This suggests that this account has been highly involved in the discourse before it is suspended by Twitter. 

Studying such users' practices and networks can help to design active measures to efficiently counter such campaigns. It can also help the posterior analysis of such users' effect on the rest of the discourse. The impact of such studies will go beyond the scope of the country and will help general research in this area.

\subsection{Natural Language Processing}
Natural Language Processing (NLP) research benefits from large textual datasets. Research on several languages is limited by the availability of such large datasets. We note that the most common language in this data is Urdu, with over 34 million tweets and over 1 million tweets in Roman Urdu. Roman Urdu can also help the research on Roman Hindi, and help differentiate both effectively. As we note that many tweets are identified as Hindi, that is because of the large overlap of vocabulary between the two languages. These tweets provide a rich data source that can be used to train language models for tasks such as sentiment analysis and generating embedding-based models.

\section{Conclusion}
In this paper, we shared the largest Twitter dataset related to Pakistani political discourse. The dataset is particularly based on April 2022 political crisis. We also highlight several use cases of this data such as identifying political bias, studying political engineering, and natural language process research. These use cases can help the research related to Pakistani Twittersphere and contribute to social network analysis research, in general.




\bibliographystyle{ieeetr}
\bibliography{references,ref}

\end{document}